\documentstyle[aps,twocolumn,epsf]{revtex}
\begin{document}
\draft
\title{Micromagnetic Simulation of nanoscale Films with perpendicular
  Anisotropy}
\author{U.~Nowak}
\address{ 
  Theoretische Tieftemperaturphysik\\ 
  Gerhard-Mercator-Universit\"{a}t-Duisburg\\
  47048 Duisburg/ Germany\\
  e-mail: uli@thp.uni-duisburg.de}
\date{\today}
\maketitle
\begin{abstract}
  A model is studied for the theoretical description of nanoscale
  magnetic films with high perpendicular anisotropy. In the model the
  magnetic film is described in terms of single domain magnetic grains
  with Ising-like behavior, interacting via exchange as well as via
  dipolar forces. Additionally, the model contains an energy barrier
  and a coupling to an external magnetic field. Disorder is taken into
  account in order to describe realistic domain and domain wall
  structures. The influence of a finite temperature as well as the
  dynamics can be modeled by a Monte Carlo simulation.

  Many of the experimental findings can be investigated and at least
  partly understood by the model introduced above. For thin films the
  magnetisation reversal is driven by domain wall motion. The results
  for the field and temperature dependence of the domain wall velocity
  suggest that for thin films hysteresis can be described as a
  depinning transition of the domain walls rounded by thermal
  activation for finite temperatures.
\end{abstract}

\section{Introduction}
We focus on thin ferromagnetic films with high perpendicular
anisotropy like $CoPt$ and $MnBi$ which can be used as magneto-optical
storage media. In these films two different mechanisms can be thought
to dominate the reversal process: either nucleation or domain wall
motion \cite{Pommier}. Which of these mechanisms dominates a reversal
process depends on the interplay of the different interaction forces
between domains with different magnetic orientation. In a recent
experiment on $Co_{28}Pt_{72}$ alloy films \cite{Theo,Valentin} a
crossover from magnetisation reversal dominated by domain growth to a
reversal dominated by a continuous nucleation of domains was found
depending on the film thickness which was varied from 100{\AA} to
300{\AA}.  Correspondingly, characteristic differences for the
hysteresis loops have been found. Similar results have been achieved
by simulations of a micromagnetic model using energy minimization
techniques \cite{Theo,Valentin} and Monte Carlo methods, respectively
\cite{Nowak}.  It is the goal of this paper to investigate the
influence of the field and the temperature on the domain wall velocity
for the case of magnetisation reversal dominated by domain wall motion
by a Monte Carlo simulation of a micromagnetic model.

\section{Micromagnetic Model}
$Co_{28}Pt_{72}$ alloy films have a polycrystalline structure with
grain diameters of 100-250{\AA}. For a theoretical description by a
micromagnetic model \cite{Andra} the film is thought to consist of
cells on a square lattice with a square base of size $L \times L$
where $L = 200${\AA} and height $h$ of 100{{\AA}}.  Due to the high
anisotropy of the $Co_{28}Pt_{72}$ alloy film the grains are thought
to be magnetised perpendicular to the film only with a uniform
magnetisation $M_s$ which is set to the experimental value of $M_s =
365$kA/m for the saturation magnetisation in these systems
\cite{Aachen}. The grains interact via domain wall energy and dipole
interaction. The coupling of the magnetisation to an external magnetic
field $H$ is taken into account as well as an energy barrier which has
to be overcome during the reversal process of a single cell.

From these considerations it follows that the change of energy caused
by reversal of a cell $i$ with magnetisation $L^2 h M_s \sigma_i$ with
$\sigma_i = \pm 1$ is:
\begin{eqnarray}
\Delta E_i &=& -\frac{1}{2} L h S_w \Delta \sigma_i \sum_{<j>} \sigma_j
             +\frac{\mu_0}{4\pi} M_s^2 L h^2 \Delta \sigma_i
                             \sum_j v(\sigma_j,r_{ij}) \nonumber\\
           &&  -\mu_0 H L^2 h M_s \Delta \sigma_i
\label{e.ham}
\end{eqnarray}

The first term describes the wall energy $\Delta E_w$. The sum is over
the four next neighbors and for $S_w$ we use a value of $S_w = 0.0022
\mbox{J}/\mbox{m}^2$ which is approximately 50\% of the Bloch-wall
energy $S_B$ for this system \cite{Aachen}. The reason for this
reduction of the grain interaction energy compared to the Bloch-wall
energy is, that the crystalline structure of the system is interrupted
at the grain boundary and also that due to their irregular shape the
grains are not connected via their complete surface $Lh$.

In the second term describing the dipole coupling $\Delta E_d$ the sum
is over all cells. $r_{ij}$ is the distance between two cells $i$ and
$j$ in units of the lattice constant $L$. For large distances it is
$v(\sigma_j,r_{ij}) = \frac{\sigma_j}{r_{ij}^3}$. For shorter
distances a more complicated form which is a better approximation
for the shape of the cells which we consider can be determined
numerically and was taken into account.

The third term describes the coupling $\Delta E_H$ to an external
field $H$.

Additionally, an energy barrier $\delta_i$ must be considered which
describes the fact that a certain energy is needed to reverse an
isolated cell by domain wall motion through the grain (see also
\cite{Ruediger}). We assume that during the reversal process the
energy barrier has its maximum value $L h S_b$ when the domain wall is
in the center of the cell, i.e. when half of the cell is already reversed.
Consequently, the energy barrier which is relevant for the reversal
process is reduced to $\delta = \max(0, L h S_b - \frac{1}{2} |(\Delta
E_w + \Delta E_d + \Delta E_H)|)$. The simulations are in good
qualitative agreement with experiments using $S_b = 0.0007
\mbox{J}/\mbox{m}^2$.

In order to simulate the $CoPt$ films realistically disorder has to be
considered. Obviously, the grain sizes are randomly distributed
\cite{Theo}. In the model above this corresponds to a random
distribution of $L$ which can hardly be simulated exactly since it
modulates the normalized cell distance $r_{ij}$ of the dipole
interaction. Therefore, as a simplified ansatz to simulate the
influence of disorder we randomly distribute $L$ in the energy term
that describes the coupling to the external field. Here a random
fluctuation of $L$ is most relevant, since this term is the only one
scaling quadratic ly with $L$. In the simulations we use a
distribution which is Gaussian with width $\Delta$. Note, that through
this kind of disorder our model is mapped on a random-field model.

The simulation of the model above was done as in an earlier
publications \cite{Nowak,Ruediger} via Monte Carlo methods
\cite{Binder} using the Metropolis algorithm with an additional energy
barrier. Since the algorithm satisfies detailed balance and Glauber
dynamics it allows the investigation of thermal properties as well as
the investigation of the dynamics of the system. For temperatures $T
\rightarrow 0$ the Monte Carlo algorithm passes into a simple energy
minimization algorithm with single spin flip dynamics, so that also the
case of zero temperature can be investigated.

The size of the lattice was typically $150 \times 150$. The dipole
interaction was taken into account without any cut-off or mean field
approximation.

\section{Hysteresis and Dynamics}

Fig.~\ref{f.hys} shows a simulated hysteresis loop.  The loop is
nearly rectangular.  Here, the reversal is dominated by domain wall
motion \cite{Nowak}. Once a nucleus begins to grow the domain wall
motion does not stop until the magnetisation has completely changed.
The nucleation field for the simulation is too high compared to the
corresponding experimental results \cite{Theo,Valentin}.  There are
several possible reasons for this effect. First, the nucleation field
which in the case of domain wall motion dominated reversal is the
field where domain wall motion starts depends on the size and on the
shape of the nucleus. In an experimental situation a nucleus can be
very large e.~g. a scratch while there is no such artificial nucleus
in our simulation except of the disorder which leads only to nuclei of
very small size. However, it is not the aim of our simulations to
calculate the nucleation field accurately.  Rather, the simulations
are thought to contribute to a better understanding of the fundamental
properties of the system.

Since the reversal is driven by domain wall motion, the velocity of
the domain wall is a central quantity which we will investigate in the
following.  For the determination of the domain wall velocity within
the simulation we start with a system that has a nucleus of circular
shape with a radius of 19 cells in the center of the $150 \times 150$
system. When we switch on the driving field, from the nucleus a domain
starts to grow. For the better observation of the domain growth, in
our flip-algorithm we do not consider cells that are not connected to
the growing domain, i.~e. we exclude the possibility of additional
nucleation. Otherwise we have -- at least for finite temperatures --
the problem that spontaneously new nuclei are build by thermal
activation which with increasing radius overlap with the original
domain.  Fig.~\ref{f.dom} shows a domain during the reversal. The
black region is the reversed domain following the magnetic field. It
has a circular shape with a rough domain wall.  From the domain
configurations the mean radius $r$ of the domains can be determined
through the area $F$ of the reversed domain as $r = \sqrt{F/\pi}$,
assuming that the domain has a circular shape.

In order to get a deeper understanding of the influence of the
temperature on the dynamics we simulated the reversal for temperatures
$T=0, 300,$ and 600K.  Fig.~\ref{f.kt0} shows - as an example - the
$r(t)$ behavior from the simulations for $T=0$K and different fields.
For $r > 75$ the domain reaches the boundary of the system and -
consequently - $r(t)$ saturates. For zero temperature the domain wall
is pinned for lower fields, i.~e. after a short period of
rearrangement of the domain wall the domain wall movement stops and
the radius remains constant.  The pinning of the domain wall is due to
energy barriers which follow from the disorder, the dipole field, and
the intrinsic energy barrier of the single cell. For finite
temperatures, the domain wall velocity is always finite.

For $20 < r < 75$ the slope of the $r(t)$ curve is approximately
constant and $v$ can be determined by fitting to a straight line.
Fig.~\ref{f.vh} shows the dependence of the domain wall velocity on
the driving field for $T=0, 300$, and 600K. For zero temperature there
is a sharp depinning transition \cite{Kardar} at a critical field
$H_c$ from a pinned phase with $v = 0$ to a phase with finite domain
wall velocity. This transition can be interpreted in terms of a phase
transition with $v \sim (H-H_c)^\theta$ for $H > H_c$ where in our
case the critical exponent is $\theta \approx 1$, a value which is the
mean field result for a moving elastic interface \cite{Leschhorn} in a
random field. Also, this value has been observed in simulations of a
soft spin model with random-fields \cite{Jost}.  For finite
temperatures the transition is smeared since for finite temperatures
there is even for $H<H_c$ for each energy barrier a finite probability
that the barrier can be overcome by thermal fluctuations. The
corresponding waiting time can be expected to be exponentially large
so that for $H < H_c$ the domain wall velocity should decrease like
$\ln v \sim (H-H_c)/T$.  To illustrate this in Fig.\ref{f.skal} we
show the corresponding semi-logarithmic scaling plot. As we expect,
the data for the two different finite temperatures collapse for
$H<H_c$ on a straight line.  For $H>H_c$ thermal activation is
obviously less relevant. Here, the dynamics is dominated by the zero
temperature depinning transition.

To conclude, the results for the domain wall velocity suggest that for
zero temperature the hysteresis driven by domain wall motion can be
understood as a depinning transition of the domain walls. For finite
temperature the transition is rounded and for fields smaller than the
depinning field the domain wall movement is dominated by thermal
activation. A paper on a comparison of these theoretical results with
experimental measurements of the domain wall velocity in $CoPt$ alloy
films is in preparation. 

\acknowledgments{The author thanks K.~D.~Usadel for helpful
  discussions and for critically reading the manuscript.}

\begin{figure}
  \caption{Hysteresis loop for a 100{\AA} film; $T=300$K}
  \label{f.hys}
\end{figure}

\begin{figure}
  \caption{Domain configuration of a $150 \times 150$ system during
  the reversal; $T=300$K}
  \label{f.dom}
\end{figure}

\begin{figure}
  \caption{Radius of the domain versus time for the same
    fields as in Fig.~\ref{f.vh}; $T=0$; solid lines are best fitted.}
  \label{f.kt0}
\end{figure}

\begin{figure}
  \caption{Domain wall velocity versus driving field for $T = 0,
    300$, and 600K; solid lines are guides to the eye.}
  \label{f.vh}
\end{figure}

\begin{figure}
  \caption{
  Scaling plot from Fig.~\ref{f.vh}.}
  \label{f.skal}
\end{figure}


\begin{references}
\bibitem{Pommier} J.~Pommier, P.~Meyer, G.~P\'{e}nissard,
  J.~Ferr\'{e}, P.~Bruno, and D.~Renard, Phys.~Rev.~Lett. {\bf 65},
  2054 (1990)
\bibitem{Theo} T.~Kleinefeld, J.~Valentin, D.~Weller, JMMM {\bf 148},
  249 (1994)
\bibitem{Valentin} J.~Valentin, T.~Kleinefeld, D.~Weller,
  J.~Phys.~D {\bf 29}, 1111 (1996)
\bibitem{Nowak} U.~Nowak, IEEE Trans.~Mag.~{\bf 31}, 4169 (1995)
\bibitem{Andra} W.~Andr\"{a}, H.~Danan, and R.~Mattheis,
  Phys.~Stat.~Sol. (a), {\bf 125} , 9 (1991)
\bibitem{Aachen} J.~Harzer, RWTH Aachen, Ergebnisbericht (1992)
\bibitem{Ruediger} U.~Nowak, U.~Ruediger, P.~Fumagalli,
  G.~G\"{u}ntherodt, Phys.~Rev.~B {\bf 54}, 13017 (1996)
\bibitem{Binder} K.~Binder, Monte Carlo Methods in Statistical
  Physics (Springer-Verlag, Berlin 1979)
\bibitem{Kardar} M.~Kardar and D.~Ertas, in Scale Invariance,
  Interfaces, and Non-Equilibrium Dynamics, edited by A.~McKane,
  M.~Droz, J.~Vannimenus, and D.~Wolf, Plenum Press, New York 1995,
  pa. 89
\bibitem{Leschhorn} H.~Leschhorn, J.~Magn.~Magn.~Mat. {\bf 104-107},
  309 (1992)
\bibitem{Jost} K.~D.~Usadel and M.~Jost, J.~Phys. A{\bf 26}, 1783 (1993)
\end{references}
\end{document}